\documentclass[english,twocolumn,twocolumn,showpacs,preprintnumbers,amsmath,amssymb,superscriptaddress]{revtex4}
\usepackage{graphicx}

\makeatletter

\providecommand{\tabularnewline}{\\}

\makeatletter

\makeatletter

\makeatletter

\usepackage{color}

\usepackage{dcolumn}

\usepackage{bm}

\newcommand{\pt}{ p_{\rm t}}
\newcommand{\ie}{{\it i.e.}}

 \def\lsim{\mathrel{\rlap{\lower4pt\hbox{\hskip1pt$\sim$}}
    \raise1pt\hbox{$<$}}}         %less than or approx. symbol
 \def\gsim{\mathrel{\rlap{\lower4pt\hbox{\hskip1pt$\sim$}}
    \raise1pt\hbox{$>$}}}         %greater than or approx. symbol

\makeatother

\usepackage{babel}
\makeatother
\begin{document}

\title{Centrality-dependent Direct Photon $\pt$ spectra in Au+Au Collisions
at RHIC}

\author{Fu-Ming Liu}

\email{liufm@iopp.ccnu.edu.cn}

\affiliation{Institute of Particle Physics, Central China Normal University, Wuhan,
China }

\author{Tetsufumi Hirano}

\affiliation{Department of Physics, The University of Tokyo, 113-0033, Japan}

\author{Klaus Werner}

\affiliation{Laboratoire SUBATECH, University of Nantes - IN2P3/CNRS - Ecole desMines,
Nantes, France }

\author{Yan Zhu}

\affiliation{Institute of Particle Physics, Central China Normal University, Wuhan,
China }

\date{\today}

\begin{abstract}
We calculate the centrality-dependence of transverse momentum ($\pt$)
spectra for direct photons in Au+Au collisions at the RHIC energy,
based on a realistic data-constrained (3+1) dimensional hydrodynamic
description of the expanding hot and dense matter, a reasonable treatment
of the propagation of partons and their energy loss in the fluid,
and a systematic study of the main sources of direct photons. The
resultant $\pt$ spectra agree with recent PHENIX data in a broad
$\pt$ range. The competition among the different direct photon sources
is investigated at various centralities.
Parton energy loss in the plasma is considered for
photons from fragmentation and jet photon conversion, which 
causes about 40\% decrease in the total contribution. 
In the high $\pt$  region, the observed $R_{AA}$ of photons is centrality 
independent at the accuracy of 5\% based on a realistic treatment of 
energy loss. 
We
also link the different behavior of $R_{AA}$ for central and peripheral
collisions, in the low $\pt$ region, to the fact that the plasma
in central collisions is hotter compared to peripheral ones. 
\end{abstract}
\maketitle
%\pacs{}
%\keywords{}

\section{Introduction}

The formation and observation of a quark-gluon plasma (QGP) in heavy
ion collisions are important goals of modern nuclear physics~\cite{QM06,harris96}.
Suppression of high $\pt$ hadron yields~\cite{supression} is one
of the most important features observed at the Relativistic Heavy
Ion Collider (RHIC). Theoretically this is attributed to the interaction
between jets (hard partons) and the bulk matter \cite{bjorken82,Gy90,Baier97,BDMPS}.
Experimentally, absence of the suppression in $d+$Au collisions~\cite{dAu}
reveals that the suppression results from a final-state effect and,
in turn, that the hot and dense matter is created in Au+Au collisions.
The amount of suppression depends significantly on the centrality
of the collision \cite{AuAu-pi-phenix}, which implies various sizes
of hot dense matter are formed in heavy ion collisions at various
centralities. This offers us an excellent opportunity to study the
interaction of partons inside the system and, consequently, properties
of the matter under extreme conditions.

Hadron production in heavy ion collisions involves bulk hadronization
of the thermal partons at low $\pt$, the fragmentation of quenched
hard partons at high $\pt$ and the hadronization contributed from
both thermal and hard partons at intermediate $\pt$. However, it
is quite difficult to systematically describe all these hadronization
processes since some of them are beyond the perturbative treatment
and usually contain many parameters without full understanding. Low
$\pt$ hadrons also strongly interact with each other after hadronization
and cannot carry \textit{direct} information from inside the hot matter.
Under this situation, a systematic study of direct photons in a wide
range of transverse momentum and centrality can serve as a guide to
understand the whole reaction processes of heavy ion collisions, since
we do not need to treat hadronization itself nor interaction between
produced direct photons and the bulk matter thanks to the large mean
free path of direct photons compared to the typical size of the system
in heavy ion collisions. Competition among different sources at various
centralities may be also useful in understanding of production mechanism
of direct photons.

In this paper, we first study the role of jet quenching on the centrality
dependence of direct photon production. For this purpose, a reliable
treatment of hard parton energy loss is needed. This is formulated
via the BDMPS framework \cite{BDMPS} and tested on pion suppression
at various centralities. Since neutral pions and other mesons are
significantly suppressed in central Au+Au collisions, and since the
suppression has an evident centrality dependence, the following question
arises naturally: What is the role of hard parton energy loss on direct
photon production? Main purpose of this paper is to answer this question.

We also investigate the interplay among the various sources of direct
photons. Similar to hadron suppression, photons from parton fragmentation
are expected to offer information on the interaction between hard
partons and the bulk via jet quenching. Thermal photons and photons
from parton-bulk interactions are penetrating probes of the hot matter,
respectively, through interaction of partons inside bulk matter and
interaction between primary partons and the bulk matter. It is interesting
to see whether one reproduces the observed photon spectra by considering
all photon sources simultaneously and consistently at different collision
centralities. By identifying the dominant sources of direct photons
at given values of $\pt$, we will be able to discuss in which way
the different $\pt$-regions of the photon spectra provide information
about the different production processes.

We need a realistic description of the hot and dense matter to investigate
the effect of bulk matter on photon emission. This is achieved by
using three dimensional (3D) hydrodynamic simulations of bulk matter
\cite{Hirano:2001eu,Hirano02} which have already been tested against
a vast body of low $\pt$ hadron data at RHIC.

To perform a systematic study of \textit{direct} photon production
(from sources other than neutral meson decays) in relativistic heavy
ion collisions, we shortly review the possible sources in the following.

\textbf{\textit{Primordial NN scattering.}} The direct photon production
via Compton scattering and quark-antiquark annihilation can be calculated
in perturbation theory using the conventional parton distribution
functions and the factorization hypothesis. In principle one should
consider at this stage also higher order contributions, like bremsstrahlung
of photons accompanying for example two-jet production in hard parton-parton
scattering. However, we consider this component as a part of the so-called
jet fragmentation (or bremsstrahlung) contribution to be affected
by the thermalized matter, which we will discuss separately.

\textbf{\textit{Thermal photons.}} In high energy nuclear collisions,
the density of secondary partons is so high that the quarks and gluons
rescatter and eventually thermalize to form a bubble of hot QGP. The
plasma expands, cools down, and goes through a phase transition to
hadronic gas (HG) phase. Thermal photons can be produced during the
whole history of the evolution of hot matter from the QGP phase to
the HG phase through the mixed phase due to collisions of or radiations
from thermalized particles. Yields of photons from a thermal source
are exponentially damped so that contribution to very high $\pt$
region is negligible. However, contribution to low $\pt$ is expected
to be dominant in central collisions in which the size and the temperature
of a hot and dense matter are large enough.

\textbf{\textit{Jet-photon conversion.}} When hard partons pass through
thermalized matter, they may interact. Collisions between jets and
deconfined partons via quark-antiquark annihilation and quark-gluon
Compton scattering can produce direct photons. This is often called
as jet-photon conversion.

\textbf{\textit{Jet fragmentation.}} Photon production also occurs
as a higher order effect in purely partonic initial hard scatterings:
at any stages of the evolution of a jet (final state parton emission),
there is a possibility of emitting photons. Existence of a QGP again
affects the results of fragmented photons since energetic partons
lose their energy prior to fragmentation. In this work, we assume
fragmentation of partons only outside the plasma, which is similar
to high $\pt$ hadron production from jet fragmentation.

There are possible contributions to photon production which are not
included in the present study: The medium-induced radiation is supposed
to be a higher order contribution. At the RHIC energy this contributes
much less than jet photon conversion at low and intermediate $\pt$
and much less than fragmentation at high $\pt$~\cite{Turbide:2005fk}.
So the contribution from medium-induced radiation is ignored this
paper. In the time interval between the primordial collisions at $\tau=0$
and the thermalization of the hot matter at $\tau_{0}$, the interaction
between non-equilibrated soft partons and hard partons may also produce
direct photons. We neglect the contribution in the preequilibrium
stage since the time interval is much shorter than the life time of
the equilibrated matter ($\sim20$ fm/$c$).

The paper is organized as follows: In Sec.~\ref{sec:2}, we first
give a brief review on the space-time evolution of the hot matter
created in Au+Au collisions at different centralities based on a (3+1)-dimensional
ideal hydrodynamical calculation. In Sec.~\ref{sec:3}, we discuss
parton energy loss in the QGP. We investigate neutral pion production
in the high $\pt$ region in order to fix the parameters of the energy
loss scheme. We discuss sequently the contributions from various sources
to direct photon $\pt$ spectra in Sec.~\ref{sec:4}. We show our
results and compare them with recent experimental data in Sec.~\ref{sec:5}.
Section~\ref{sec:6} is devoted to conclusion of the present study.

\section{space-time evolution of the hot and dense matter }

\label{sec:2}

Several sources of direct photon production in heavy ion collisions
depend on the bulk dynamics of hot and/or dense matter and the matter
along trajectories of energetic partons. So a realistic description
of reaction dynamics is indispensable for the quantitative analysis
of photon production. In our calculation, fully three-dimensional
(3D) ideal hydrodynamics \cite{Hirano:2001eu,Hirano02} is employed
to describe the space-time evolution of the hot and dense matter created
in Au+Au collisions at RHIC energy at various centralities. We solve
the equations of energy-momentum conservation \begin{eqnarray}
\partial_{\mu}T^{\mu\nu}=0\end{eqnarray}
 in full 3D space $(\tau,x,y,\eta)$ under the assumption that the
local thermal equilibrium is reached (maintained) at (after) an initial
time $\tau_{0}$ =0.6 fm/$c$. Here $\tau$ and $\eta$ are the proper
time and the space-time rapidity, respectively. $x$ and $y$ are
transverse coordinates. In the transverse plane, the centers of two
colliding nuclei are located at $(x,y)=(b/2,0)$ and $(-b/2,0)$ before
the collision at an impact parameter $b$. Ideal hydrodynamics is
characterized by the energy-momentum tensor, \begin{equation}
T^{\mu\nu}=(e+P)u^{\mu}u^{\nu}-Pg^{\mu\nu},\end{equation}
 where $e$, $P$, and $u^{\mu}$ are energy density, pressure, and
local four velocity, respectively. We neglect the finite net-baryon
density which is small near the mid-rapidity at RHIC. For the high
temperature ($T>T_{c}=170$ MeV) QGP phase we use the equation of
state (EOS) of massless non-interacting parton gas ($u$, $d$, $s$
quarks and gluons) with a bag pressure $B$: \begin{eqnarray}
p=\frac{1}{3}(e-4B).\end{eqnarray}
 The bag constant is tuned to be $B^{\frac{1}{4}}=247.19$\,MeV to
match pressure of the QGP phase to that of a hadron resonance gas
at critical temperature $T_{c}=170$\,MeV. A hadron resonance gas
model at $T<T_{c}$ includes all hadrons up to the mass of the $\Delta(1232)$
resonance. Our hadron resonance gas EOS implements chemical freeze-out
at $T_{\mathrm{ch}}=T_{c}=170$\,MeV, as observed in collisions at
RHIC \cite{BMRS01}. This is achieved by introducing appropriate
temperature-dependent chemical potentials $\mu_{i}(T)$ for all hadronic
species $i$ in a way that their numbers $\tilde{N}_{i}$ including
all decay contributions from higher-lying resonances, $\tilde{N}_{i}=N_{i}+\sum_{R}b_{R\rightarrow iX}N_{R}$,
are conserved during the evolution \cite{Hirano02,Bebie,spherio,Teaney:2002aj,Kolb:2002ve,Huovinen:2007xh}.
Here $N_{i}$ is the average multiplicity of the $i$-th hadron species,
and $b_{R\rightarrow iX}$ is the effective branching ratio (a product
of branching ratio and degeneracy) of a decay process $R\rightarrow i+X$.
In this partial chemical equilibrium (PCE) model \cite{Hirano02}
only strongly interacting resonances with large decay widths (whose
decays do not alter $\tilde{N}_{i}$) remain chemically equilibrated
below $T_{\mathrm{ch}}$. It should be noted that the hadronic chemical
composition described by hydrodynamics using the PCE model is roughly
consistent with that of the hadronic cascade models \cite{Hirano06},
as long as the latter are initialized at $T_{\mathrm{sw}}=169$ MeV
with thermal and chemical equilibrium distributions.

We assume that, at $\tau_{0}=0.6$ fm/$c$, the initial entropy distributions
is proportional to a linear combination of the number density of participants
(85\%) and binary collisions (15\%) \cite{Hirano06}. Centrality
dependence of charged particle multiplicity observed by PHOBOS \cite{PHOBOS_Nch}
has been well reproduced by full 3D hydrodynamics simulations with
the above setups~\cite{Hirano06}. In the following calculations,
hydrodynamic outputs at representative impact parameters $b=$ 3.2,
5.5, 7.2, 8.5, 9.7, and 10.8 fm are chosen for 0-10\%, 10-20\%, $\cdots$,
50-60\% centrality, respectively.

So far, the space-time evolution of the QGP fluid obtained as above
has been also exploited for a quantitative study of hard and rare
probes such as azimuthal jet anisotropy, nuclear modification factor
of identified hadrons, disappearance of back-to-back jet correlation,
and $J/\psi$ suppression~\cite{HiranoCollectPapers}.%
\begin{table}

\caption{Initial temperature at the plasma center at initial time $\tau_{0}=0.6$
fm/$c$ for various centralities.}

\begin{tabular}{|c|c|c|c|c|c|c|}
\hline 
Centrality(\%)&
0-10 &
10-20 &
20-30 &
30-40 &
40-50 &
50-60 \tabularnewline
\hline 
$T_{0}$(MeV) &
370&
357&
341&
327&
301&
272\tabularnewline
\hline
\end{tabular}\label{table:initialT} 
\end{table}

\begin{figure}
\includegraphics[scale=0.85]{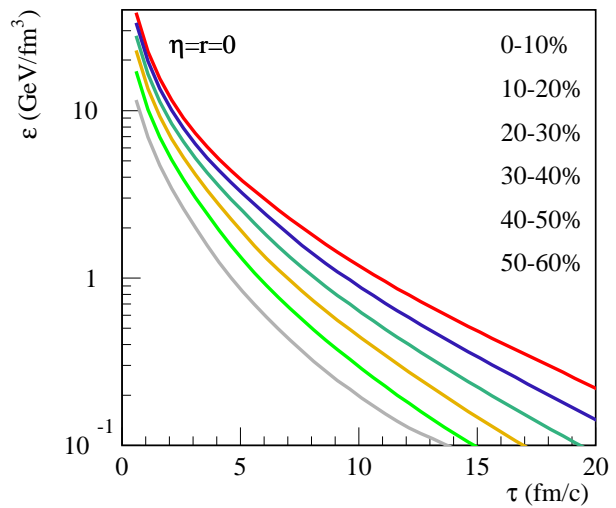}

\caption{\label{fig:veps} (Color Online) Time evolution of energy density
at the center $(x,y,\eta)=(0,0,0)$ for various centrality. Each line
from top to bottom corresponds to 0-10\%, 10-20\%, ..., and 50-60\%
centrality, respectively. }
\end{figure}

In Table \ref{table:initialT}, initial temperatures at the plasma
center, $T_{0}=T(\tau_{0},0,0,0)$, are shown for various centralities.
These temperature values will be important to interpret the centrality
dependence of the slope of pt spectra from thermal radiation, which
will be discussed later. Figure \ref{fig:veps} shows the time evolution
of energy density at the center of fluids $(x,y,\eta)=(0,0,0)$ for
various centralities. Clearly for any given proper time $\tau$, the
more central collisions one obtains higher energy densities at the
plasma center.

For convenience of the following calculations, we introduce $f_{{\rm \textrm{QGP}}}(\tau,x,y,\eta)$
as the fraction of the QGP phase in a fluid element. It is obvious
that $f_{{\rm \textrm{QGP}}}=1$ (0) in the QGP (hadronic) phase.
In the mixed phase, the fraction of the QGP is calculated via \begin{eqnarray*}
f_{\textrm{QGP}}\, e_{\textrm{QGP}}+(1-f_{\textrm{QGP}})e_{\textrm{had}}=e(\tau,x,y,\eta)\end{eqnarray*}
 with $e_{\textrm{QGP}}$ and $e_{\textrm{had}}$ being the energy
densities of the QGP phase and the hadron phase at $T=T_{c}$, respectively.

\section{Parton energy loss in a plasma}

\label{sec:3}

Energy loss of hard partons in a plasma affects both jet photon conversion
and jet fragmentation. The momentum distribution of jets (energetic
gluons or quarks with different flavors) from primordial nucleus-nucleus
scattering is calculated as~\cite{Owens1987} \begin{eqnarray}
 &  & \frac{dN^{AB\rightarrow{\rm jet}}}{dyd^{2}\pt}=KT_{AB}(b)\sum_{abcd}\int dx_{a}dx_{b}G_{a/A}(x_{a},M^{2})\nonumber \\
 &  & \qquad\times G_{b/B}(x_{b},M^{2})\frac{\hat{s}}{\pi}\frac{d\sigma}{d\hat{t}}(ab\rightarrow cd)\delta(\hat{s}+\hat{t}+\hat{u})\label{eq:ABtojet}\end{eqnarray}
 where $T_{AB}(b)$ is the nuclear overlapping function at an impact
parameter $b$ for each centrality, $G_{a/A}(x_{a},M^{2})$ and $G_{b/B}(x_{b},M^{2})$
are parton distribution functions in nuclei $A$ and $B$. We take
MRST 2001 LO parton distributions in proton~\cite{MRST0201}. The
elementary cross sections for $ab\rightarrow cd$ can be found in
Ref.~\cite{Owens1987}. We set the factorization scale $M$ and
renormalization scale $Q$ to be $M=Q=\pt$. $K=2$ is chosen to take
into account higher order contributions. These parameters are chosen
as to reproduce high $\pt$ pion data in $pp$ collisions at RHIC,
which will be discussed later. The above formula for $\pt$ spectra
was extensively tested in $pp$ ($p\bar{p}$) collisions in an energy
range from $\sqrt{s}=27.4$~GeV to 630~GeV. Nuclear shadowing effect
and EMC effect are taken into account through EKS98 scale dependent
nuclear ratios $R_{a}^{{\rm \textrm{EKS}}}(x,A)$~\cite{EKS98}.
Isospin of a nucleus with mass $A$, neutron number $N$, and proton
number $Z$ is corrected as follows: \begin{equation}
G_{a/A}(x)=\left[\frac{N}{A}G_{a/N}(x)+\frac{Z}{A}G_{a/P}(x)\right]R_{a}^{{\rm \textrm{EKS}}}(x,A).\label{eq:PdisA}\end{equation}
 The isospin mixture and nuclear shadowing eventually cause a decrease
of nuclear modification at high $\pt$ region, which will be shown
in Sec.~\ref{sec:5}.

We assume that all jets are produced at $\tau=1/Q\approx0$ with the
phase space distribution \begin{equation}
f_{0}(\vec{p},\vec{r})\propto\frac{dN}{d^{3}p}T_{A}\left(x-\frac{b}{2},y\right)T_{B}\left(x+\frac{b}{2},y\right)\delta(z)\label{eq:space dis}\end{equation}
 where $\vec{r}=(x,y,z)$ is the coordinate of a jet, $b$ is the
impact parameter, and $T_{A}$ and $T_{B}$ are thickness functions
of nuclei $A$ and $B$. The $\delta$-function reflects the highly
Lorentz-contracted colliding nuclei $A$ and $B$. The phase space
distribution of hard partons is normalized as \begin{equation}
\int f_{0}(\vec{p},\vec{r})d^{3}r=(2\pi)^{3}\frac{dN}{d^{3}p}\label{eq:fqi}\end{equation}

Energetic partons can suffer interactions with the fluid and lose
their energies. We employ the BDMPS formula~\cite{Baier97} to calculate
parton energy loss in a plasma created in heavy ion collisions. For
a parton of type $i=q,g$ with initial momentum $\vec{p}_{0}$ formed
at $\vec{r}_{0}$, the whole path length of a parton traversing the
expanding QGP (including the mixed phase) is \begin{equation}
L(\vec{p}_{0},\vec{r}_{0})=\int_{\tau_{0}}^{\infty}d\tau\,\theta\big(f_{{\rm QGP}}(\tau,\bm{x}(\tau))\big).\label{eq:L}\end{equation}
 Here $\bm{x}(\tau)$ is a trajectory of a parton, $f_{{\rm QGP}}(\tau,\bm{x}(\tau))$
is the fraction of the QGP phase at a position $(\tau,$$\bm{x}(\tau))$,
and $\theta$ is a step function, which gives $\theta(f_{{\rm QGP}})$
equal unity in the QGP and the mixed phases and zero in the hadron
phase.

The total energy loss along this path is calculated as \begin{equation}
\Delta E(i,\vec{p}_{0},\vec{r}_{0})=D\int_{\tau_{0}}^{\infty}d\tau\epsilon(i,\tau,\bm{x}(\tau))\,\theta\big(f_{{\rm QGP}}(\tau,\bm{x}(\tau))\big).\label{eq:BDMPS}\end{equation}
 Here $D$ is an adjustable parameter, $\epsilon(i,\tau,\bm{x}(\tau))$
is the energy loss per unit distance for a parton $i$ at a position
$(\tau,$$\bm{x}(\tau))$, given as~\cite{Baier97} \[
\epsilon(i,\tau,\bm{x}(\tau))=\alpha_{s}\sqrt{\mu^{2}E^{*}/\lambda_{i}}.\]
 Here, the temperature-dependent running coupling constant -- assuming
a similar formula as the lowest order one in perturbation theory --
can be obtained by fitting the numerical results from lattice quantum
chromodynamics (QCD) simulations~\cite{Karsch1998} as \begin{equation}
\alpha_{s}(T)=\frac{6\pi}{(33-2N_{f})\ln(8T/T_{c})}.\label{eq:alphasT}\end{equation}
 The Debye screening mass is given as $\mu=gT$, with $g^{2}/4\pi=\alpha_{s}(T)$.
The energy of a hard parton in the local rest frame is $E^{*}=p^{\mu}u_{\mu}$
where $p^{\mu}$ is the four momentum of the hard parton in the laboratory
frame and $u_{\mu}$ is a local fluid velocity. All hard partons are
treated as on-shell massless particles. The mean free path, $\lambda_{i}$,
of a parton $i$, is given as \begin{eqnarray}
\lambda_{g}^{-1} & = & \sigma_{gq}\rho_{q}f_{{\rm QGP}}+\sigma_{gg}\rho_{g}f_{{\rm QGP}},\label{eq:labdag}\\
\lambda_{q}^{-1} & = & \sigma_{qq}\rho_{q}f_{{\rm QGP}}+\sigma_{qg}\rho_{g}f_{{\rm QGP}},\end{eqnarray}
 with cross sections $\sigma_{i}=C_{i}\alpha_{s}\pi/T^{2}$~\cite{GW94}.
The color factors $2C_{i}$ are 4/9, 1, and 9/4 for $qq$, $qg$,
and $gg$ scattering, respectively. The parton densities $\rho_{q}$
and $\rho_{g}$ are obtained from the EOS of the massless relativistic
gas. The fraction of the QGP phase $f_{{\rm QGP}}$ is considered
in the mixed phase to ensure a smooth transition from the QGP phase
to the HG phase. Note that the above quantities, $\ie$, temperature
$T$, fluid velocity $u_{\mu}$, parton densities $\rho_{i}$ and,
in turn, mean free path $\lambda_{i}$, depend on the location of
the parton $\bm{x}(\tau)$ and can be obtained from full 3D hydrodynamics
simulations discussed in the previous section.

Various sizes of the plasma are formed in heavy ion collisions at
different centralities. We use the common energy loss formula Eq.~(\ref{eq:BDMPS})
for all of these media. The main purpose in the present paper is a
systematic study of direct photon production rather than a detailed
analysis of parton energy loss. So we admit ourselves to introduce
an adjustable parameter $D$ in Eq.~(\ref{eq:BDMPS}) to fit simultaneously
neutral $\pi$-meson data in Au+Au collisions at different centralities~\cite{AuAu-pi-phenix}.

\begin{figure}
\includegraphics[scale=0.85]{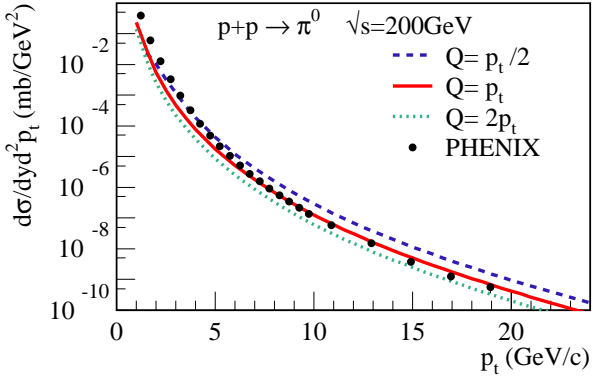}

\caption{\label{fig:pp-pi0} (Color Online) Neutral pion production in pp
collisions at $\sqrt{s}=200$ GeV is compared to PHENIX data~\cite{pp-pi-phenix}.
Three lines from top to bottom correspond to $Q=\pt/2$, $\pt$, and
$2\pt$ respectively. }
\end{figure}

We first discuss pion production in proton-proton collisions. We calculate
neutral $\pi$-meson production assuming pQCD factorization, Eq.~(\ref{eq:ABtojet}),\begin{equation}
\frac{dN_{pp}^{\mathrm{\pi^{0}}}}{dyd^{2}\pt}=\sum_{c=g,q_{i}}\int dz_{c}\frac{dN^{pp\rightarrow c}}{dyd^{2}\pt^{c}}\frac{1}{z_{c}^{2}}D_{\pi^{0}/c}^{0}(z_{c},Q^{2}),\label{eq:jet-pi0}\end{equation}
 where $D_{\pi^{0}/c}^{0}(z_{c},Q^{2})$ is $\pi^{0}$ fragmentation
functions parameterized by Kniehl \textit{et al.}~\cite{KKPfrag}.
In Fig.~\ref{fig:pp-pi0}, $\pt$ spectra for neutral pions in $pp$
collisions at $\sqrt{s}=200$ GeV calculated with $M=Q=2\pt$, $\pt$,
and $\pt/2$ are compared to PHENIX data~\cite{pp-pi-phenix}. In
the high $\pt$ region where the pQCD is expected to work, we reasonably
reproduce the experimental data with the above setup with $K=2$ and
$M=Q=\pt$. We use the $\pt$ spectrum as a reference spectrum in
the following calculations.

\begin{figure}
\includegraphics[scale=0.75]{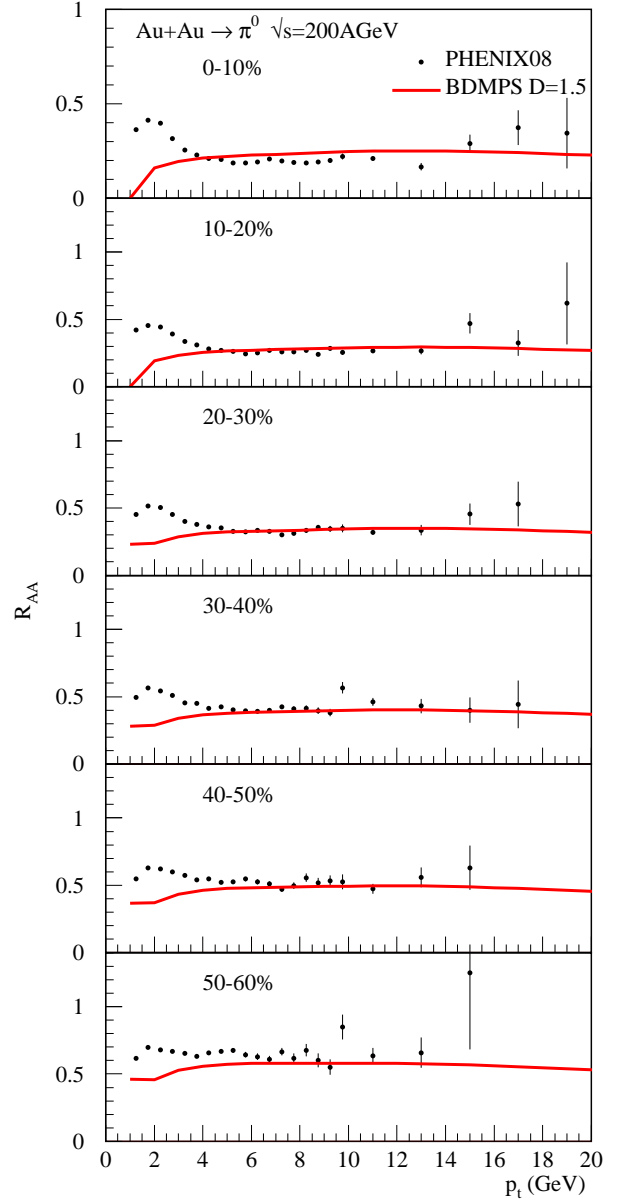}

\caption{\label{fig:RaapB} (Color Online) Nuclear modification factors of
$\pi^{0}$. Solid lines are calculated with the BDMPS energy loss
formula with amplified parameter $D=1.5$ (See text for details).
Plots are PHENIX data~\cite{AuAu-pi-phenix}.}
\end{figure}

The effect of parton energy loss is taken into account through the
medium modified fragmentation function $D_{\pi^{0}/c}(z_{c},Q^{2},\Delta E_{c})$
which describes suppression of neutral pion yields as \begin{equation}
\frac{dN_{AB}^{\pi^{0}}}{dyd^{2}\pt}=\sum_{c=g,q_{i}}\int dz_{c}\frac{dN^{AB\rightarrow c}}{dyd^{2}\pt^{c}}\frac{1}{z_{c}^{2}}D_{\pi^{0}/c}(z_{c},Q^{2},\Delta E_{c}).\label{eq:jet-pi02}\end{equation}
 with \cite{XNWANG04} \begin{eqnarray}
 &  & D_{\pi^{0}/c}(z_{c},Q^{2},\Delta E_{c})\nonumber \\
 &  & =\left(1-e^{-\frac{L}{\lambda_{c}}}\right)\left[\frac{z_{c}'}{z_{c}}D_{\pi^{0}/c}^{0}(z_{c}',Q^{2})+\frac{L}{\lambda_{c}}\frac{z_{g}'}{z_{c}}D_{\pi^{0}/g}^{0}(z_{g}',Q^{2})\right]\nonumber \\
 &  & \quad+e^{-\frac{L}{\lambda_{c}}}\: D_{\pi^{0}/c}^{0}(z_{c},Q^{2}).\end{eqnarray}
 Figure \ref{fig:RaapB} shows the nuclear modification factors for
neutral pions in Au+Au collision at $\sqrt{s_{NN}}=200$ GeV for different
centralities. Solid lines are results with an energy loss parameter
$D=1.5$ and plots are the PHENIX data~\cite{AuAu-pi-phenix}. With
a common value of the parameter $D=1.5$, we can reasonably reproduce
the $\pi^{0}$ yields in the high $\pt$ region at all centralities
simultaneously. It should be noted that, in PHENIX data~\cite{AuAu-pi-phenix},
there are $\sim10$\% normalization error and $\sim7-16$\% errors
(depending on centrality) due to $N_{\mathrm{coll}}$, which are omitted
in Fig.~\ref{fig:RaapB}. In the region $\pt<5$GeV/c, our results
undershoot the experimental data due to absence of neutral pion production
from bulk components in this calculation. Notice that low $\pt$ pion
data have already been described well \cite{Hirano06} by using hydrodynamic
simulations employed in the present study. In the following photon
calculations, we always use the BDMPS energy loss formula (\ref{eq:BDMPS})
with $D=1.5$.%

\section{The different sources of direct photon production}

\label{sec:4}

\emph{Leading order contribution.} Similar to Eq.~(\ref{eq:ABtojet}),
the leading order contribution to direct photon production in nucleus-nucleus
collisions reads \begin{eqnarray}
 &  & \frac{dN^{AB\rightarrow{\rm \gamma}}}{dyd^{2}\pt}=T_{AB}(b)\sum_{{\displaystyle ab}}\int dx_{a}dx_{b}G_{a/A}(x_{a},M^{2})\nonumber \\
 &  & \times G_{b/B}(x_{b},M^{2})\frac{\hat{s}}{\pi}\frac{d\sigma}{d\hat{t}}(ab\rightarrow\gamma+X)\delta(\hat{s}+\hat{t}+\hat{u})\label{eq:ABtogamma}\end{eqnarray}
 where the elementary processes $ab\rightarrow\gamma+X$ are Compton
scattering $qg\rightarrow\gamma q$ and annihilation $q\bar{q}\rightarrow g\gamma$.

\emph{Fragmentation} \emph{contribution.} Higher order contributions
in $pp$ collisions are due to jet fragmentation. We can calculate
them as \begin{equation}
\frac{dN_{pp}^{\mathrm{frag}}}{dyd^{2}\pt}=\sum_{c=g,q_{i}}\int dz_{c}\frac{dN^{pp\rightarrow c}}{dyd^{2}\pt^{c}}\frac{1}{z_{c}^{2}}D_{\gamma/c}^{0}(z_{c},Q^{2}),\label{eq:jet-gamma0}\end{equation}
 with photon fragmentation functions $D_{\gamma/c}^{0}(z,Q^{2})$
being the probability for obtaining a photon from a parton $c$ which
carries a fraction $z$ of the parton's momentum. So $\pt^{c}=\pt/z_{c}$
is the transverse momentum carried by the parton $c$ before fragmentation
and $d^{3}p/E=z_{c}^{2}d^{3}p^{c}/E^{c}$. The effective fragmentation
functions for obtaining photons from partons can be calculated perturbatively.
We use the parameterized solutions by Owens~\cite{Owens1987}.

In case of heavy ion collisions, parton energy loss in a plasma should
be taken into account. This can be done via modified fragmentation
functions~\cite{XNWANG04} \begin{eqnarray}
 &  & D_{\gamma/c}(z_{c},Q^{2},\Delta E_{c})\nonumber \\
 &  & =\left(1-e^{-\frac{L}{\lambda_{c}}}\right)\left[\frac{z_{c}'}{z_{c}}D_{\gamma/c}^{0}(z_{c}',Q^{2})+\frac{L}{\lambda_{c}}\frac{z_{g}'}{z_{c}}D_{\gamma/g}^{0}(z_{g}',Q^{2})\right]\nonumber \\
 &  & \quad+e^{-\frac{L}{\lambda_{c}}}\: D_{\gamma/c}^{0}(z_{c},Q^{2}),\end{eqnarray}
 with $z_{c}'=\pt/(\pt^{c}-\Delta E_{c})$ and $z_{g}'=(L/\lambda_{c})\,\pt/\Delta E_{c}$
being the rescaled momentum fractions carried by the parton $c$ and
the emitted gluons before fragmentation. % Here $D_{\gamma/q}^{0}$
% and $D_{\gamma/g}^{0}$ are the fragmentation functions in vacuum~\cite{Owens1987}.
$\lambda_{c}$ is mean free path of the parton $c$ in the plasma,
$L$ is the path length of each parton traversing the plasma defined
in Eq.~(\ref{eq:L}). Thus, in heavy ion collisions, contributions
from fragmentation become \begin{equation}
\frac{dN_{AB}^{{\rm frag}}}{dyd^{2}\pt}=\sum_{c=g,q_{i}}\int dz_{c}\frac{dN^{AB\rightarrow c}}{dyd^{2}\pt^{c}}\frac{1}{z_{c}^{2}}D_{\gamma/c}(z_{c},Q^{2},\Delta E_{c}).\label{eq:jet-gamma2}\end{equation}

The above formula counts only fragmented photons outside the plasma.
In principle, when fragmentation into photons happens inside the plasma
photon can escape the plasma due to the long mean free path. However
it is not evident when and where fragmentation happens.

\emph{Thermal production.} The emission rate of photons is $\Gamma=Ed^{3}R/d^{3}p$,
where $R$ is the number of photons emitted from a medium per unit
space-time volume with temperature $T$. Total yields of thermal photons
can be obtained by summing the emission rate over the space-time volume
as \begin{equation}
\frac{dN^{{\rm thermal}}}{dyd^{2}\pt}=\int d^{4}x\Gamma(E^{*},T)\label{eq:E*a}\end{equation}
 with $d^{4}x=\tau d\tau dxdyd\eta$ and $E^{*}=p^{\mu}u_{\mu}$ being
the photon energy in the local rest frame. Here, $p^{\mu}=(\pt\cosh y,\pt\cos\phi,\pt\sin\phi,\pt\sinh y)$
is the photon's four momentum in the laboratory frame and $u_{\mu}$
is a local fluid velocity. In our calculations, the thermal photon
emission rate covers both contributions from the QGP phase and the
hadronic phase \begin{eqnarray}
\Gamma(E^{*},T) & = & f_{{\rm QGP}}\Gamma^{{\rm QGP}\rightarrow\gamma}(E^{*},T)\nonumber \\
 &  & +(1-f_{{\rm QGP}})\Gamma^{{\rm HG\rightarrow\gamma}}(E^{*},T),\label{eq:rate2parts}\end{eqnarray}
 where $f_{{\rm QGP}}$ and $T$ are the fraction of the QGP phase
and temperature of the fluid, respectively, both being obtained in
the hydrodynamic simulations. In the above formula, we calculate thermal
photon production above the thermal freeze-out temperature $T_{\mathrm{th}}=100$
MeV. The photon emission rate from $2\rightarrow2$ processes between
thermal partons, $\ie$, the QCD Compton process $qg\rightarrow\gamma q$
and annihilation $q\bar{q}\rightarrow g\gamma$, was first calculated
with the hard thermal loop resummation technique~\cite{Kapusta1991,Baier:1991em}.
Later, Landau-Pomeranchuk-Migdal (LPM) interference effect for emitted
photons turned out to be important \cite{AMY2001}, leading to \begin{eqnarray}
\Gamma^{{\rm QGP}\rightarrow\gamma}(E^{*},T) & = & \sum_{i=1}^{N_{f}}\left(\frac{e_{i}}{e}\right)^{2}\frac{\alpha\alpha_{S}}{2\pi^{2}}T^{2}\frac{1}{e^{x}+1}\nonumber \\
 &  & \times\big[\ln\left(\frac{\sqrt{3}}{g}\right)+\frac{1}{2}\ln(2x)+C_{22}(x)\nonumber \\
 &  & +C_{\mathrm{brems}}(x)+C_{\mathrm{ann}}(x)\big],\end{eqnarray}
 with $x=E^{*}/T$ and \begin{equation}
C_{22}(x)=\frac{0.041}{x}-0.3615+1.01{\rm e}^{-1.35x},\label{eq:c22}\end{equation}
 \begin{eqnarray*}
 &  & C_{\mathrm{brems}}(x)+C_{\mathrm{ann}}(x)\\
 &  & =0.633x^{-1.5}\ln\left(12.28+1/x\right)+\frac{0.154x}{(1+x/16.27)^{0.5}}\ .\end{eqnarray*}
 In the calculation we take $N_{f}=3$ and a temperature dependent
running coupling as in Eq.~(\ref{eq:alphasT}).

Thermal photon emission in the hadronic phase results from interactions
such as $\pi\pi\rightarrow\rho\gamma$, $\pi\rho\rightarrow\pi\gamma$
and $\rho\rightarrow\pi\pi\gamma$, etc. Interactions of mesons or
baryons with strangeness can also produce photons, but the contributions
are relatively small due to the phase space suppression resulting
from their heavier masses. In our work, photon emission rate from
the hadronic phase is based on massive Yang-Mills (MYM) calculation
\cite{Rapp2004}, where photon production from mesons with strangeness
has been included as well as the axial meson $a_{1}$ as an exchanging
particle for non-strange initial states. Hadrons are composite objects,
so form factors are considered to simulate finite hadronic size effects~\cite{Arleo2004}.

\emph{Jet photon conversion with jet energy loss}. When hard partons
propagate in a plasma, they also collide with thermal partons and
produce direct photons via Compton process and the quark-antiquark
annihilation. We call this process jet-photon conversion, since it
is a conversion of a jet into a photon with almost the same momentum
as the one of originating jet parton. Contribution from the jet-photon
conversion is calculated by integration of conversion rate over the
space-time evolution of the hot and dense matter in the QGP phase
\begin{equation}
\frac{dN^{{\rm jpc}}}{dyd^{2}\pt}=\int\Gamma^{{\rm jpc}}(E^{*},T)f_{{\rm QGP}}(x,y,\eta,\tau)d^{4}x.\label{eq:jpc_int}\end{equation}
 The photon production rate by annihilation and Compton scattering
of hard partons in the medium can be approximated as \cite{WangCY,Fries2005}
\begin{equation}
\Gamma^{{\rm jpc}}(E^{*},T)=\frac{\alpha\alpha_{s}}{4\pi^{2}}\sum_{q}e_{q}^{2}f_{q}(\vec{p},x)T^{2}\left[\ln\frac{4E_{\gamma}^{*}T}{m_{{\rm th}}^{2}}-C\right]\label{eq:jpc_rate}\end{equation}
 where $E^{*}$ is the photon energy in the local rest frame, $C=2.323$,
$m_{{\rm th}}^{2}=g^{2}T^{2}/6$, and the strong coupling $\alpha_{s}=g^{2}/4\pi$
being temperature dependent as in Eq.~(\ref{eq:alphasT}). $\alpha$
is the electromagnetic couplings, $e_{q}$ and $f_{q}(\vec{p},x)$
are the electric charge and the phase-space density of a hard parton
of flavor $q$. The phase space distribution of hard partons at $\tau$
is obtained by considering parton energy loss as \begin{eqnarray*}
f(\vec{p},x) & = & f(\vec{p},\vec{r},\tau)\\
 & = & \int d^{3}p_{0}f_{0}(\vec{p}_{0},\vec{r}-\vec{v}t)\delta(\vec{p_{0}}-\vec{p}-\vec{v}\Delta E)\end{eqnarray*}
 where $f_{0}(\vec{p},\vec{r})$ is the phase space distribution at
$\tau=0$ described in Eqs.~(\ref{eq:space dis}) and (\ref{eq:fqi}).
The $\delta$-function expression reflects the energy loss of a parton
moving along a straight line trajectory with $\vec{v}\equiv\vec{p}/E=\vec{p}_{0}/E_{0}$.
$\Delta E$ is the energy loss from $\tau_{0}$ to $\tau$ and calculated
similar to Eq.~(\ref{eq:BDMPS}) but replacing the upper limit of
integral $\infty$ with $\tau$.

\section{results and discussion}

\label{sec:5}

\begin{figure*}
\includegraphics[scale=0.8]{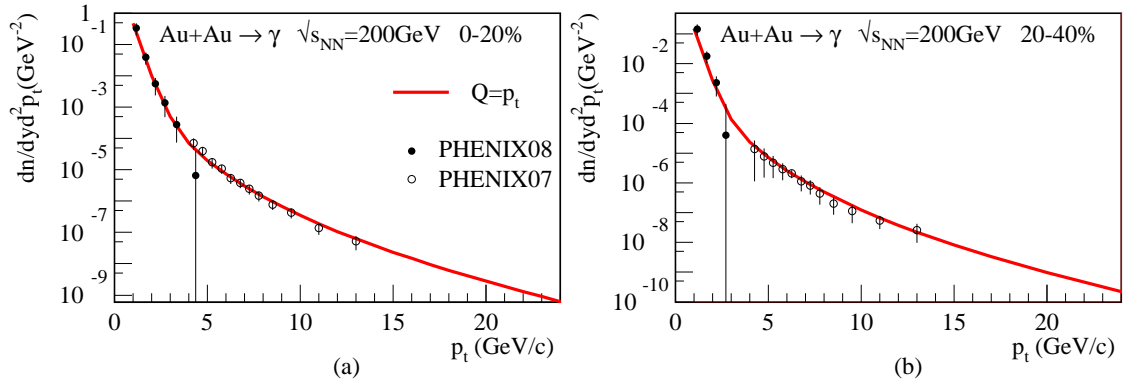}

\caption{\label{fig:c12-34} (Color Online) Direct photon production in Au+Au
collisions at centrality 0-20\% and 20-40\%. PHENIX data are shown
as open circles~\cite{PHENIX07Phys. Rev. Lett. } and filled circles~\cite{PHENIX08g}.}
\end{figure*}

\begin{figure}
\includegraphics[scale=0.85]{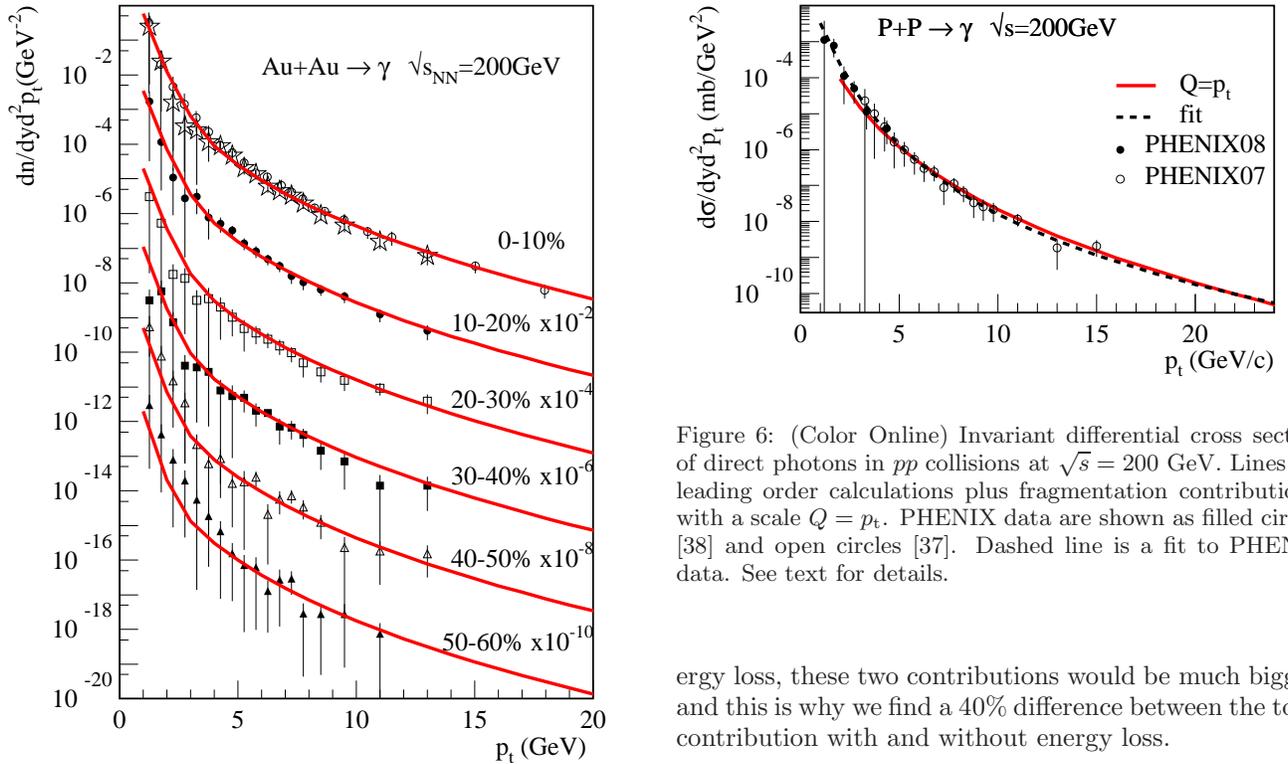}

\caption{\label{fig:fig1} (Color Online) Direct photon production in Au+Au
collisions at $\sqrt{s_{NN}}=$200 GeV for different centralities
(0-10\%, 10-20\%, ..., 50-60\%). Data are obtained by PHENIX~\cite{PHENIX data}. }
\end{figure}

In Fig.~\ref{fig:c12-34}, the calculated $\pt$ spectra of direct
photons in Au+Au collisions at $\sqrt{s_{NN}}=200$ GeV at centrality
0-20\% and 20-40\% are compared to PHENIX data \cite{PHENIX07Phys. Rev. Lett. ,PHENIX08g}.
Here we sum over all contributions discussed in the previous section.
The theoretical results for 0-20\% centrality are obtained as a mixture
of the calculations for 0-10\% and 10-20\% centrality with a weight
of 50\% each; a corresponding procedure applies for the 20-40\% centrality
results. The PHENIX data are reproduced within our multi-component
model remarkably well.

In Fig.~\ref{fig:fig1}, we show a detailed comparison of the calculated
$\pt$ spectra of direct photons with PHENIX data \cite{PHENIX data}
for the centralities 0-10\%, 10-20\%, ..., 50-60\%. Again, our results
agree with data very well in a broad range of $\pt$ and centrality.

\begin{figure}
\includegraphics[scale=0.85]{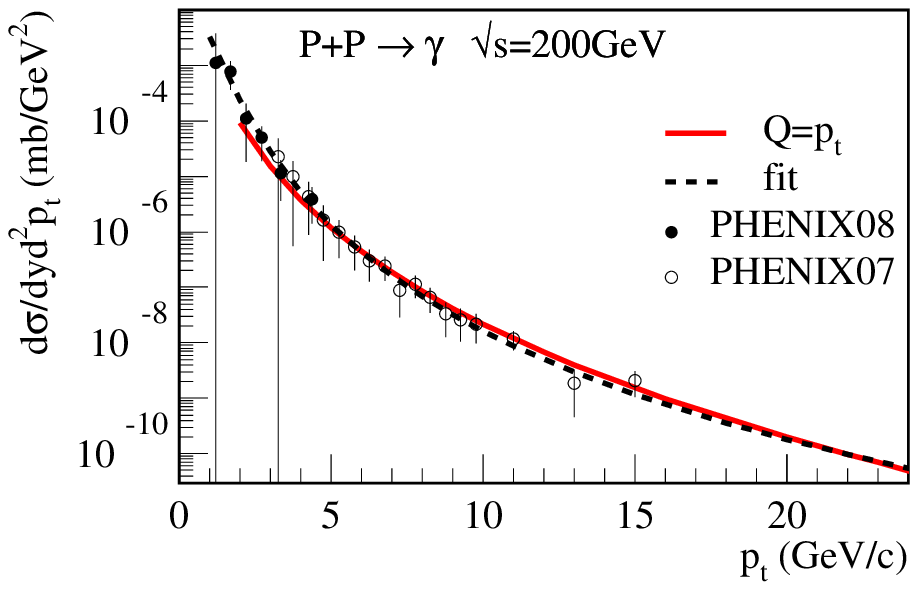}

\caption{\label{fig:pp_gamma} (Color Online) Invariant differential cross
section of direct photons in $pp$ collisions at $\sqrt{s}=200$~GeV.
Lines are leading order calculations plus fragmentation contributions,
with a scale $Q=\pt$. PHENIX data are shown as filled circles \cite{PHENIX08g}
and open circles \cite{PHENIX07Phys. Rev. Lett. }. Dashed line is
a fit to PHENIX data. See text for details. }
\end{figure}

\begin{figure*}
\includegraphics[scale=0.8]{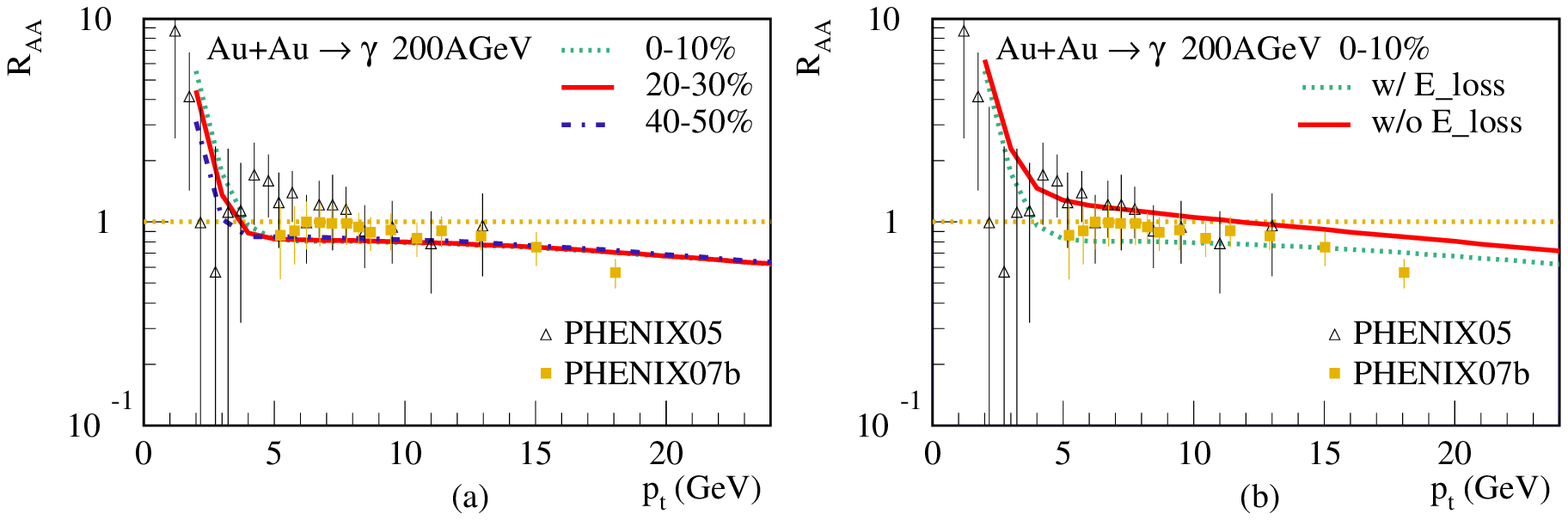}

\caption{\label{fig:Raa3} (Color Online) The nuclear modification factor
of direct photons in Au+Au collisions $R_{AA}$. Data for 0-10\% centrality
are from Ref.~\cite{PHENIX data} and Ref.~\cite{PHENIX07}. (a):
$R_{AA}$ at centrality 0-10\% (dotted line), 20-30\% (solid line),
and 40-50\% (dash-dotted line) respectively. (b): $R_{AA}$ at
0-10\% centrality with energy loss (dotted line) and without energy
loss (solid line).}
\end{figure*}

Since all the curves are almost parallel to each other, one gets more
insight by using the nuclear modification factor $R_{AA}$, obtained
by dividing a $\pt$ spectrum in nucleus-nucleus collisions by the
$N_{\mathrm{coll}}$-scaled $\pt$ spectrum in $pp$ collisions. In
Fig.~\ref{fig:pp_gamma}, we show the invariant differential cross
section of direct photons in $pp$ collisions. The calculation includes
the leading order contribution plus fragmentation contribution, using
a scale $Q=\pt$. PHENIX data are shown as open circles \cite{PHENIX07Phys. Rev. Lett. }
and filled circles \cite{PHENIX08g}. In high $\pt$ regions, our
result agrees with the data reasonably well: So we use it to calculate
nuclear modification in Fig. \ref{fig:Raa3} and Fig.~\ref{fig:Eloss2}.
It also provides a baseline calculation with the LO contribution and
fragmentation contribution in Au+Au collisions. Whereas, in low $\pt$
regions where pQCD is not expected to work, our result undershoots
the data slightly although the error bars are large in data. The dashed
line is a fit to the measured differential cross section of direct
photons in $pp$ collisions at the RHIC energy \[
\frac{d\sigma}{dyd^{2}\pt}=0.01834\left(1+\frac{\pt^{2}}{1.432}\right)^{-3.27}\quad{\rm mb/GeV}^{2},\]
 which is employed to calculate the nuclear modification factor from
thermal contribution in Fig.~\ref{fig:ther} (a).

Figure~\ref{fig:Raa3} shows how the nuclear modification factor for
direct photons, $R_{AA}$, depends on centrality and  on energy
loss. Data for 0-10\% centrality are taken from Refs.~\cite{PHENIX data}
and \cite{PHENIX07}. Figure~\ref{fig:Raa3}(a) shows centrality
dependence of $R_{AA}$ compared to the PHENIX data. The three curves
are respectively 0-10\% (dotted line), 20-30\% (solid line), and 40-50\%
(dash-dotted line). $R_{AA}$ has a weak centrality dependence at
high $\pt$ region. This result is consistent with the observed phenomenon~\cite{PHENIX data}
that the $\pt$-integrated (for $\pt>6$~GeV/$c$) $R_{AA}$ of direct
photons is almost independent of collision centrality. Does this imply
a very weak effect from jet quenching? Figure~\ref{fig:Raa3}(b)
answers this question (here for the most central collisions): Comparing
calculations with (dotted line) and without energy loss (solid line),
one finds a difference of up to 40\%. So the effect of parton energy
loss is quite visible in the $\pt$ range between $4$ GeV/$c$ and
more than $20$ GeV/$c$. If we would do the $R_{AA}$ calculations without 
energy loss, the difference between central and semiperipheral collisions 
would be about  20\%, wheras the complete calculation gives the same result 
for all centralities, within 5\%. We have to admit that we talk about small 
effects, requiring experimental data with relative errors of less than 5%, 
to observe the effects.  %
\begin{figure*}
\includegraphics[scale=0.7]{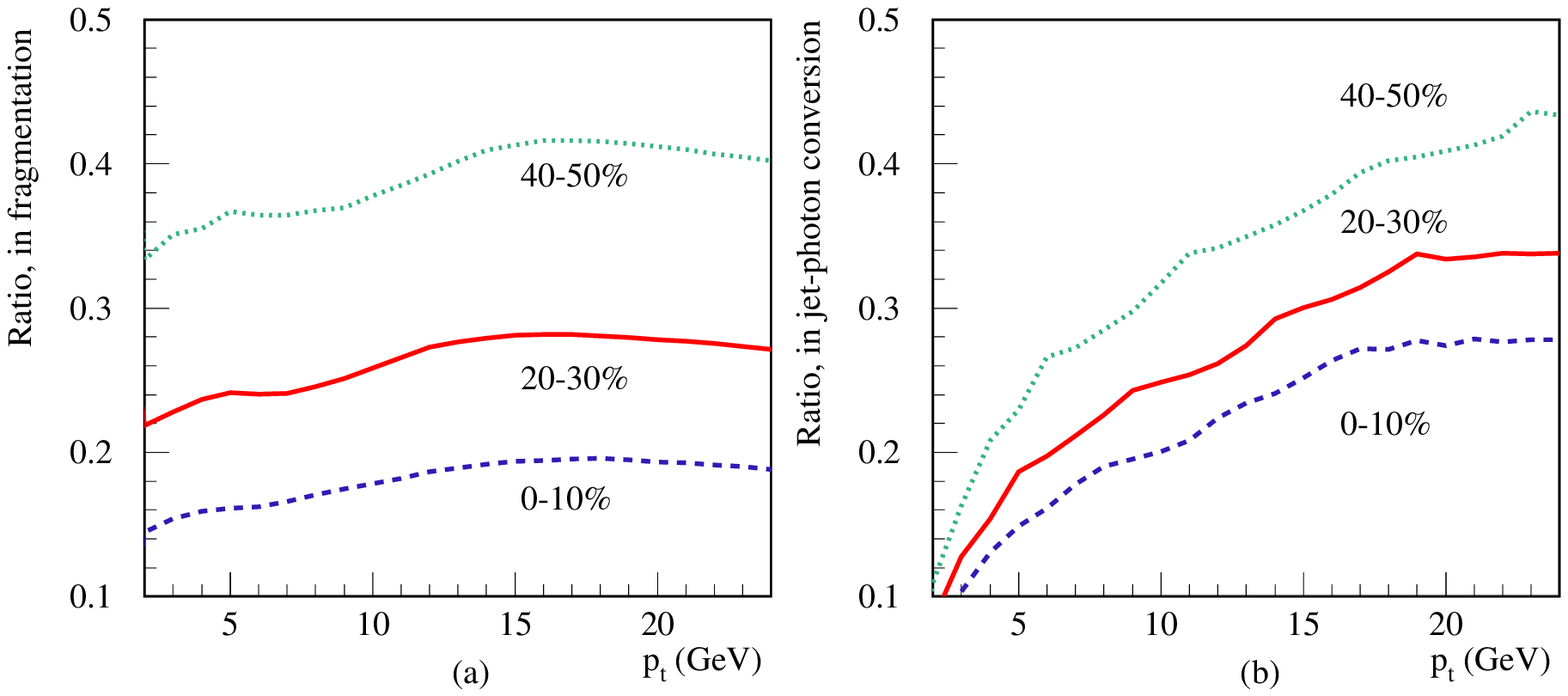}

\caption{\label{fig:Eloss} (Color Online) The ratio of the contribution with
energy loss to the one without, in fragmentation (a) and jet-photon
conversion (b).}
\end{figure*}

To understand the above results, we look more closely into the different
contributions. Parton energy loss in the plasma suppresses the fragmentation
contributions and jet-photon conversion. So we study the ratios of
the contribution with energy loss to the one without energy loss,
as shown in Fig.~\ref{fig:Eloss} ((a) for fragmentation and (b)
for jet-photon conversion). Energy loss in the plasma depends on the
path length of the hard parton inside the plasma, which turns out
to depend on the collision centrality. We do see a similar centrality
dependence of the suppression for $\pi^{0}$ (jet quenching effect)
in fragmentation contributions and jet-photon conversion.

\begin{figure*}
\includegraphics[scale=0.8]{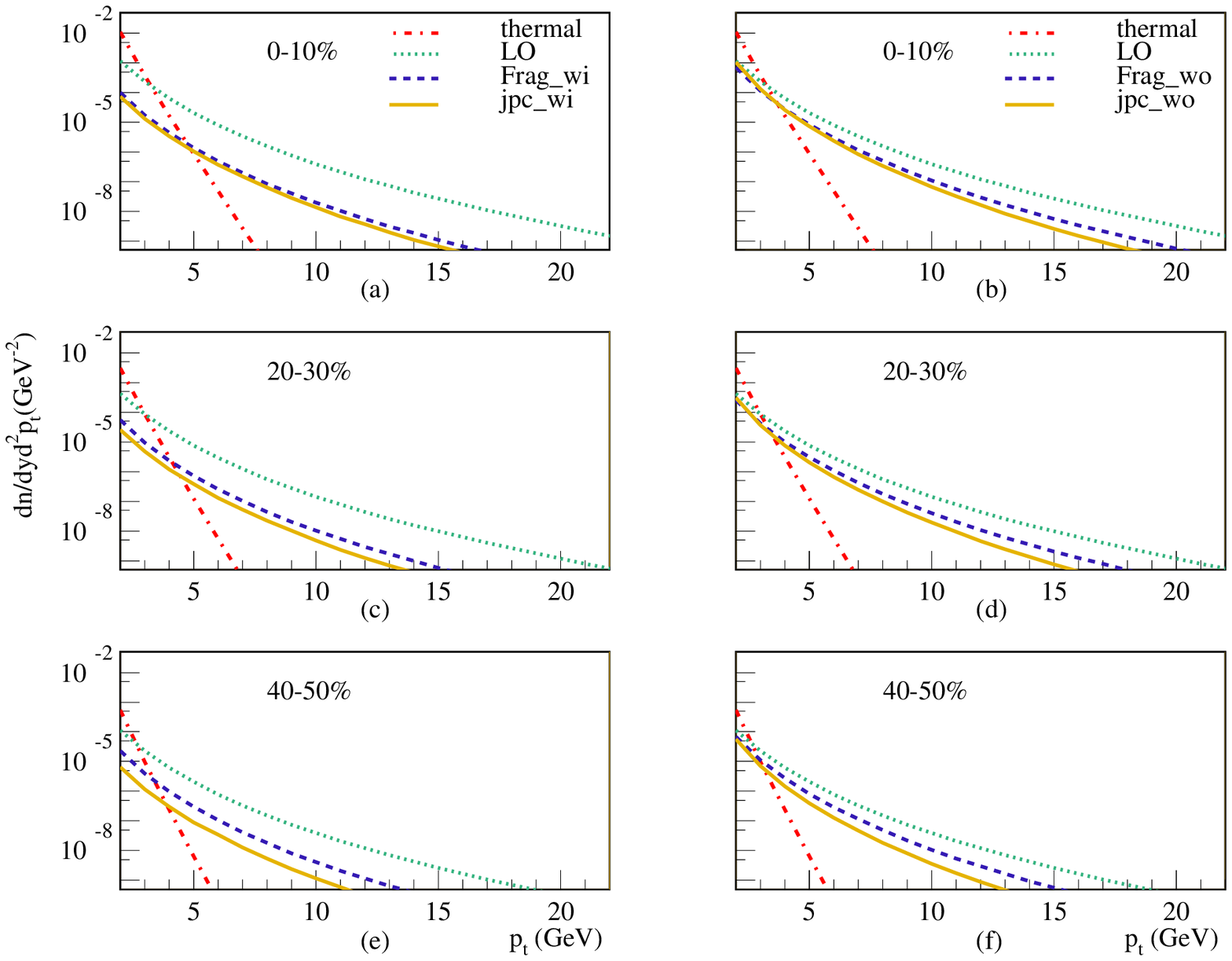}

\caption{\label{fig:cent2b} (Color Online) Competition among different sources
for direct photon production in Au+Au collisions at $\sqrt{s_{NN}}=$200~GeV
for different centralities. The leading order contribution (LO) from
primordial elementary scatterings is plotted as dotted lines, thermal
contribution as dash-dotted lines, fragmentation contribution as dashed
lines and jet-photon conversion as solid lines.}
\end{figure*}

\begin{figure*}
\includegraphics[scale=0.75]{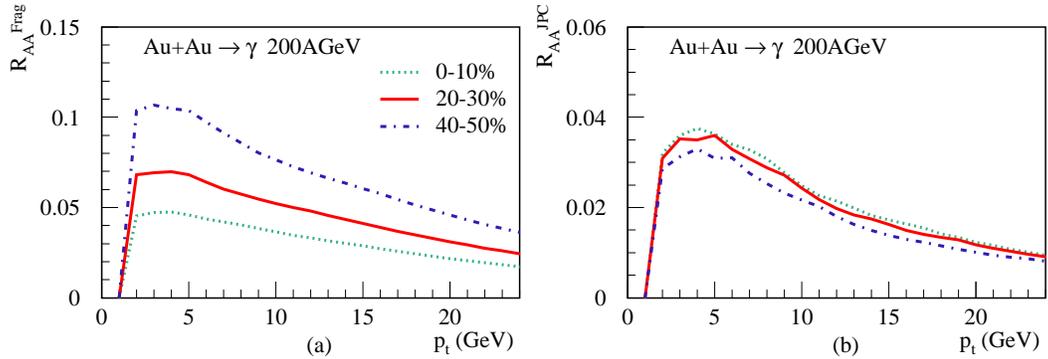}

\caption{\label{fig:Eloss2} (Color Online) The contribution to $R_{AA}$
from fragmentation and conversion, for different centralities.}
\end{figure*}

\begin{figure*}
\includegraphics[scale=0.75]{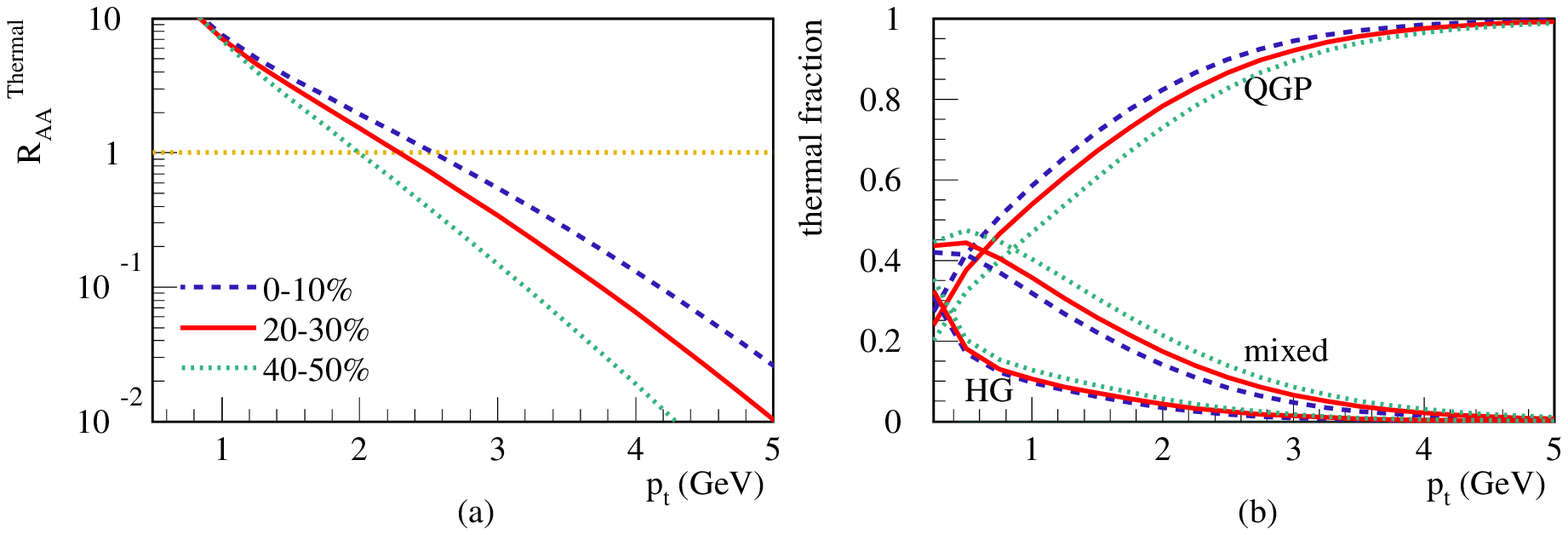}

\caption{\label{fig:ther} (Color Online) (a) Thermal contribution to $R_{AA}$
from 0-10\%~(dashed line), 20-30\%~(solid line), and 40-50\%~(dotted
line). (b) Fractions of thermal photon yields from the QGP phase,
the mixed phase, and HG phase.}
\end{figure*}

To understand how these energy loss features affect the total contribution,
we investigate the competition from different sources in Fig.~\ref{fig:cent2b},
for the three centralities 0-10\%, 20-30\%, and 40-50\%. The leading
order contribution (LO) from primordial elementary scatterings is
plotted as dotted lines, thermal contribution as dash-dotted lines,
fragmentation contribution as dashed lines, and jet photon conversion
as solid lines. The latter two are calculated with parton energy loss
in the plasma (left plots) and without (right). For all centralities,
thermal photons dominate at low transverse momenta and they are insignificant
in the high $\pt$ region. The leading order contribution from primordial
elementary scatterings dominates in the high $\pt$ region. This contribution
is independent of bulk volume.

Let us first discuss the central collisions. Here fragmentation and
conversion are an order of magnitude smaller that the LO contribution.
But from Fig.~\ref{fig:cent2b}, we know that is due to the strong
energy loss effect. Without this energy loss, these two contributions
would be much bigger, and this is why we find a 40\% difference between
the total contribution with and without energy loss.

For peripheral collisions, without energy loss, the relative contribution
from conversion is smaller compared to central scatterings, since
the the plasma regions is smaller. But then also the suppression from
energy loss is smaller for the peripheral compared to central collisions.
So at the end, including a proper energy loss treatment, for both
central and peripheral collisions, conversion is roughly an order
of magnitude smaller than the LO contribution, see Fig.~\ref{fig:cent2b}(e).
The relative contribution from fragmentation, without
energy loss, is comparable in central and peripheral collisions, however,
the energy loss is smaller in the latter ones. So fragmentation is
somewhat more important in peripheral compared to central, when energy
loss is considered, as can be also seen from Fig.~\ref{fig:Eloss2},
where the contribution to $R_{AA}$ from fragmentation and conversion
is shown for the different centralities: Conversion contributes roughly
4\%, for all centralities, fragmentation between 5\% (central) and
10\% (peripheral).

At the end, $R_{AA}$ is nearly centrality independent as shown in
Fig.~\ref{fig:Raa3}(a), but a realistic (and strong) partonic
energy loss is needed in order to get this scaling behavior.

In the low $\pt$ region, contribution from thermal radiation is of
significant importance. We check the centrality-dependence of the
thermal contribution to $R_{AA}$ in Fig.~\ref{fig:ther}~(a).
At very low $\pt$, $\ie$, $\pt<$ 1~GeV/$c$, the thermal contributions
to $R_{AA}$ at different centralities coincide with each other. However
the slope of $R_{AA}^{{\rm thermal}}$ changes and the dominant $\pt$
region of thermal photons becomes smaller when one moves from central
to peripheral collisions. This reflects the fact that the temperature
in the core region depends on the collision centrality as shown in
Table \ref{table:initialT} in Sec.~\ref{sec:2}.

So from the thermal source, the $R_{AA}$ for central collisions exceed
more and more the $R_{AA}$ for peripheral collisions, which translates
into a slight overshooting of the central total $R_{AA}$ compared
to the peripheral one, as seen in Fig.~\ref{fig:Eloss}, in the region
$\pt<$ 4~GeV/$c$.

The fractions of thermal contribution as a function of $\pt$ from
different phases are shown in Fig.~\ref{fig:ther}~(b). Partial
chemical equilibrium in the hadronic phase is used in this hydrodynamic
simulation to keep the number of hadrons fixed below $T_{\mathrm{ch}}$.
If we ignore the contribution from particle decays and use a full
chemical equilibrium (FCE), the photon emission rate from hadronic
phase \cite{Rapp2004} can be used in this case. In case of the PCE,
the chemical potential $\mu_{i}$ for all hadronic species $i$ will
modify the photon emission rate from hadronic gas, roughly estimated
by a factor of $\exp[(\mu_{1}+\mu_{2})/T]$ for a subprocess of $1+2\rightarrow3+\gamma$
according to kinetic theory with a Maxwell-Boltzmann statistics for
all particles. This factor finally increases contribution from the
HG with a factor about 2. Nevertheless, in the total thermal contribution,
PCE or FCE consideration does not make a visible difference. For all
centralities from 0-10\% to 50-60\%, the QGP phase emits most of direct
photons above $\pt\sim$1~GeV/$c$. Although the volume of hadronic
phase is much bigger than the one of the QGP phase due to expansion,
the photon emission rate from the hadronic phase decreases even faster
with temperature. The competition between volume and emission rate
results in the biggest contribution from the QGP phase at $\pt>1$~GeV/$c$.
In the current setting of hydrodynamic simulations at the RHIC energy,
the mixed phase exists for a very long time ($\sim8$ fm/$c$). This
contributes mostly at lower $\pt$ values. % For even lower $\pt$ ($\pt<gT$),
% the photon emission rate formula is not reliable because the perturbative
% expansion in the finite temperature theory is not under control. 
By combining the results shown in Figs.~\ref{fig:cent2b} and \ref{fig:ther},
contribution of thermal radiation from the QGP phase is dominant in
the region $1\lsim\pt\lsim4$ GeV/$c$. This momentum window may provide
information inside the hot and dense matter, e.g., the initial temperature
at the center, which may not be reached directly by hadron spectra.

\section{Conclusion}

\label{sec:6}

We calculated the centrality-dependence of $\pt$ spectra for direct
photons in Au+Au collisions at the RHIC energy, based on a realistic
data-constrained (3+1)-D hydrodynamic description of the expanding
hot and dense matter, a reasonable treatment of propagation of partons
and their energy loss, and a systematic consideration of main sources
of direct photons. In this study, four main sources are considered,
namely, leading order (LO) contribution from primordial elementary
scatterings, thermal radiation from the fluids, fragmentation from
hard partons, and jet photon conversion (JPC).
Similar work~\cite{Turbide:2007mi} has been done before the appearence of
the most recent data~\cite{PHENIX08g}. 
Our results agree nicely with the recent low $\pt$ data. 

The role of jet quenching in the high $\pt$ region of direct photons
production has been checked via fragmentation photons and jet photon
conversion sources. For these two sources, the suppression of the
photon rate due to parton energy loss is significant in central Au+Au
collisions, and becomes less important towards peripheral collisions,
similar to the suppression for meson production. Since experimentally
one may separate isolated photons (LO+JPC) and associate photons (fragmentation
photons), our prediction may be tested in the future.

Considering the total yields of direct photons, the contribution from
fragmentation and conversion are small, contributing between 5\% and
10\%. However, parton energy loss plays nevertheless an important
role: Without it, these second order effects would contribute significantly.
Without jet quenching, the nuclear modification factors $R_{AA}$
would depend visibly on the centrality of the collisions. A strong
energy loss is actually necessary to get the centrality scaling of
$R_{AA}$ in our calculation -- a centrality scaling which has observed
by the PHENIX collaboration. In this sense, properties of the bulk
matter affect the photon yields at intermediate values of $\pt$,
via the parton energy loss.

The low $\pt$ region is totally dominated by thermal radiation, providing
direct information about the bulk matter. We find that $R_{AA}$ of
photons at $\pt$ values below 1~GeV/$c$ is centrality independent.
With increasing $\pt$, the $R_{AA}$ for peripheral collisions drops
much faster than the one for central scatterings. On the other hand,
thermal photons are mainly emitted from the QGP phase at $\pt>$1~GeV/c
even though the mixed phase and the HG phase occupy bigger space and
longer time. So the different behavior of $R_{AA}$ for central and
peripheral collisions, in the range 1~GeV/$c<\pt<$3~GeV/$c$,
manifests the fact that the plasma in central collisions is hotter
compared to peripheral collisions.

Still more investigation are needed for a precise characterization
of the properties of the plasma via thermal photons. Besides, the
elliptic flow of direct photons (especially thermal photons) should
provide more information of the plasma, which will be discussed elsewhere.

\begin{acknowledgments}
This work is supported by the Natural Science Foundation of China
under the project No. 10505010 and MOE of China under project No.~IRT0624.
The work of T.H. was partly supported by Grant-in-Aid for Scientific
Research No.~19740130. F.M.~Liu thanks the IN2P3/CNRS and Subatech
for their hospitality during her visit in Nantes. T.H. and K.W. thank
the Institute of Particle Physics, Central China Normal University
for its hospitality during their visits. 
\end{acknowledgments}

\end{document}